\documentclass[a4paper,twocolumn,superscriptaddress]{revtex4-2}

\pdfoutput=1
\usepackage[utf8]{inputenc}
\usepackage{graphicx,bm,bbm}
\usepackage{amssymb,graphicx,epstopdf}
\usepackage{slashed,subfigure}
\usepackage{caption}
\usepackage{xcolor}
\usepackage{amsmath,mathtools}
\usepackage{epstopdf,dcolumn}
\usepackage{soul}
\usepackage{mathtools}
\allowdisplaybreaks
\usepackage[normalem]{ulem}
\usepackage{cancel}
\usepackage{xcolor}
\usepackage[colorlinks=true,linktocpage=true,linkcolor=blue,citecolor=blue]{hyperref}
\usepackage{xcolor,epsfig,amsmath,amssymb,subfigure,placeins}
\usepackage{graphicx,amsmath,bbm,bm,array}

\newcommand{\be}{\begin{equation}}
	\newcommand{\ee}{\end{equation}}
\newcommand{\ba}{\begin{eqnarray}}
	\newcommand{\ea}{\end{eqnarray}}
\newcommand{\nn}{\nonumber\\}
\newcommand{\ts}{\hspace{.3mm}}

\begin{document}
	
	\title{Li\'enard-Wiechert potential of a heavy quarkonium moving in QGP medium}
	\author{Jobin Sebastian}
	\email{jobin.sebastian@niser.ac.in}
	\affiliation{School of Physical Sciences, National Institute of Science Education and Research,
		An OCC of Homi Bhabha National Institute, Jatni-752050, India}
	\author{Mohammad Yousuf Jamal}
	\email{yousufjml5@gmail.com }
	\affiliation{School of Physical Sciences, National Institute of Science Education and Research,
		An OCC of Homi Bhabha National Institute, Jatni-752050, India}
	\affiliation{Department of Physics, B. S. Abdur Rahman Crescent Institute of Science and Technology,\\
		 Chennai 600048, India}
	 \affiliation{School of Physical Sciences, Indian Institute of Technology Goa, Ponda- 403401 Goa, India}
	\author{Najmul Haque}
	\email{nhaque@niser.ac.in}
	\affiliation{School of Physical Sciences, National Institute of Science Education and Research,
		An OCC of Homi Bhabha National Institute, Jatni-752050, India}

	\begin{abstract}
		We investigate the nature of the complex retarded potential of a heavy quarkonium moving in a hot and dense static deconfined nuclear medium. The well-known concept of the retarded potential in electrodynamics is extended to the context of the heavy-quark by modifying the static vacuum Cornell potential through Lorentz transformation to the static frame of the medium. The resulting potential in the vacuum is further corrected to incorporate the screening effect offered by the thermal medium. To do so, the retarded Cornell potential is modified by the dielectric function of the static quark-gluon plasma (QGP) medium. We present the numerical results for the real and imaginary parts of the potential along with the analytical expression of the potential approximated by a small velocity limit. 
		Finally, we present the thermal width of quarkonia in the QGP medium derived using the imaginary part of the potential and study its dependence on velocity and temperature.
	\end{abstract}

	%
	\maketitle%
	
	\bigskip
	
	\section{Introduction}
	The quark gluon-plasma (QGP) is the primordial matter at high temperature and density composed of elementary particles known as quarks, antiquarks, and gluons, free to move beyond the nucleon volume. This phase is believed to exist after the big bang when the universe was of the age of a few microseconds. During the evolution of the universe i.e., on further cooling and spatial expansion, the QGP underwent a phase transition to the hadronic matter i.e., confined state of quarks and gluons known as hadrons. Their dynamics could be governed by the laws of strong interactions where the corresponding theory is quantum chromodynamics (QCD). The advanced experimental facilities for heavy-ion collisions (HIC), such as the Relativistic Heavy Ion Collider at the Brookhaven National Laboratory and the Large Hadron Collider at CERN, provide us a unique opportunity to study the strongly interacting deconfined quark and gluon matter but within some limitations, {\it viz., } the QGP medium formed in these experiments has a very small size ($\sim 10$ fm) and very short-lived ($\sim 10^{-23}$ sec)~\cite{STAR:2005gfr, PHENIX:2004vcz, PHOBOS:2004zne, BRAHMS:2004adc}. Therefore,  the very short persistence of the medium created in HIC restricts the possibility of exploring it through external probes to quantitatively characterize its properties. Therefore, we mostly rely on internal probes to look through the created matter. In this context, the heavy quarks (charm and bottom)  and their bound states, i.e., heavy quarkonia, have a huge significance~\cite{Kurian:2020orp, Prakash:2021lwt, Prino:2016cni, Montagnoli:2017tho,Dumitru:2007hy}. They are mostly created, due to their higher masses, at the very early stages after the collisions and behave almost as an independent degree of freedom while passing through the several phases of the created matter. Though they get merely affected by the QGP medium while passing through it, resulting in distinctive signatures in their final yields observed at the detectors. In Ref.~\cite{Matsui:1986dk}, Matsui and Satz suggested that the heavy quarkonium production would be suppressed in high-energy heavy-ion collisions due to the Debye screening offered by the plasma that reduces the effective interaction between constituent particles. 
	
	In a similar scenario, we want to invite some attention toward the velocity dependence on the heavy quarkonium potential moving in the QGP medium. To study that the Cornell potential~\cite{Eichten:1974af, Eichten:1979ms} which is a linear combination of the Coulomb and linear potentials, plays an important role as it successfully takes into account the two crucial features of QCD, namely the asymptotic freedom (at a small distance or high energy) and the quark confinement (at a large distance or low energy)~\cite{Petronilo:2021usw}. The Cornell potential also played an important role in the study of various aspects of the heavy quarkonia. That includes the transition between the confined and deconfined phases of matter~\cite{Vega:2014vsa} and the calculation of the masses of various heavy quarkonium states. The authors in Ref.~\cite{Karsch:1987pv} first applied the potential models to study various quarkonia states at finite temperatures. Later, the quarkonium spectral functions and the meson current correlators have also been obtained from potential models~\cite{Mocsy:2004bv, Wong:2004zr, Mocsy:2005qw, Cabrera:2006wh, Mocsy:2007jz, Alberico:2007rg, Mocsy:2008eg, Karsch:2000gi} and are compared to the first-principle lattice QCD calculations~\cite{Asakawa:2003re, Datta:2003ww, Aarts:2007pk}. Additionally, it has also been studied that the imaginary part of the potential due to the interaction with the medium leads to the thermal dissociation width of quarkonia states~\cite{Laine:2006ns, Beraudo:2007ky}. The authors in Ref.~\cite{Jamal:2018mog, Agotiya:2016bqr} have studied the quarkonia dissociation in the anisotropic QCD medium. Several analysis on the velocity dependence of the screening properties have been carried out in Refs.~\cite{Escobedo:2011ie, Escobedo:2013tca, Liu:2006nn, Chakraborty:2006md, Caceres:2006ta}. The medium modified potential of a static quarkonium in a moving thermal bath and its velocity dependence were studied in Refs.~\cite{Thakur:2016cki} when the quark-antiquark pair is along or perpendicular to the direction of the velocity of the medium.
	
	In this article, we aim to study the heavy quarkonium potential in a context where the QGP medium is static and uniform, and the heavy quarkonia are moving with respect to the rest frame of the medium. It is similar to the situation of the retarded potential of a moving charged particle in electromagnetic plasma (EMP) or the general Li\'enard-Wiechert potential in the context of the QCD. The results obtained here for a static and uniform medium do not automatically refer to a clean suppression signal in a rapidly expanding QGP. But it serves the purpose of seeing the relative motion between a heavy quarkonia and the QGP medium that breaks the spherical symmetry of the potential. It further helps to understand the modification of the binding of quark and antiquark pairs that, in turn, modify the survival probabilities of quarkonia states observed in an asymmetric emission pattern called ``anisotropic flow." The motivation behind the current analysis is to study the effects of temperature, screening, and velocity on the retarded potential of the moving quarkonia in the static QGP medium and its angular dependence while in motion. In this article, we provide a framework to study the Li\'enard-Wiechert/ retarded potential of an open heavy flavor that will lead to quarkonium bound states potential inside the QGP. To do so, we write the Cornell potential in a covariant form and then perform a Lorentz transformation to go to the static QGP frame where the heavy quark is moving. Later, we modify this potential using the dielectric permittivity of the QGP medium.  There we observed both the real and the imaginary parts of the retarded potential.  In the current manuscript our aim is to study the chromoelectric interactions only. Incorporating the color magnetic field effects will be beyond the scope of this article. In our framework, we studied the full angular dependence of the retarded potential and showed the corresponding plots in the results section. Along with the derivation of the analytical expression within the small velocity limit, we also present the full numerical results to compare and check for the validity of the assumption. 
		Further, we use the imaginary part of the potential to calculate the thermal width of the quarkonia and study its dependence on velocity and temperature.	
		
		The paper is organized as follows. In Sec.~\ref{2}, we present the derivation of the retarded potential of a heavy quark moving in the vacuum. In Sec.~\ref{3}, we obtain the complex form of in-medium quarkonium retarded potential using the static dielectric permittivity of the QGP medium. Section ~\ref{4} is dedicated to the derivation of the potential at small velocities. In Sec.~\ref{5}, we obtain the thermal width of both charmonium and bottomonium states. Section ~\ref{6} is the results section where we discuss the velocity, temperature, and angular dependence of the medium-modified retarded potential and the thermal width. Section~\ref{7} is devoted to a summary and conclusion of the present work. 
		Natural units are used throughout the text with $c=k_B=\hbar=1$. We use a bold typeface to indicate three-vectors and a regular font to indicate four-vectors.
		The center dot depicts the four-vector scalar product with the formula $g_{\mu\nu}={\text {diag}}(1,-1,-1,-1)$.
		
		
		\section{Formalism}
		\label{2}
		Correspondence from the QED with theoretical consistency makes it easier to understand the hot QCD medium. Specifically, the QED plasma resembles QCD plasma (QGP) in some special cases, such as at the soft scale where the field fluctuation is of the order of $\sqrt{g}$ and small coupling~\cite{Blaizot:2001nr}. Here, we are employing the analogy of Li\'enard-Wiechert potential from the electrodynamics and proceeding to obtain the retarded potential for the heavy quarkonium inside the QGP medium. For that, we first consider the (static) Cornell potential that binds the quark-antiquark pair in the vacuum and write it in a covariant form. Next, we know that the four-potential in a particular frame can be transformed into any other frame using Lorentz transformations. The four-potential corresponds to heavy quark-antiquark interaction in its rest frame with the Cornell potential as the scalar part is given by
		\ba
		A^{\mu}_{0} =\left(-\frac{\alpha}{r}+\sigma r,{\bm 0} \right),\label{CORNELLREST}
		\ea
		where $r=|{\bm r}|$ is the distance from heavy quark to the field point;  $\alpha=C_F\alpha_s$  with $C_F=\left(N_c^2-1\right)/2N_c$ and  $\alpha_s$ is the strong coupling constant; $\sigma$ is the string tension; and $N_c$ is the number of color degrees of freedom. The four-potential $A^{\mu}_{0}$ in Eq.~\eqref{CORNELLREST} can be written in the covariant form by introducing the four-velocity $u^{\mu}_{0}\equiv(1,{\bm 0})$ in the rest frame of the heavy quark. So, in the rest frame of the heavy quark the four-potential is 
		\ba
		A^{\mu}_{0}=\left(-\frac{\alpha}{r_\nu u^\nu_0}+\sigma~ (r_\nu u^\nu_0),{\bm 0} \right) ,
		\label{eq:Amu}
		\ea
		Note that $r_\nu u^\nu_0=r$. Now, the Lorentz transformations of Eq.~\eqref{eq:Amu} to a frame where the heavy quark is moving with a velocity, ${\bm v}$ is given as
		\ba
		A^{\mu}\!=\!\left(-\frac{\gamma \alpha }{r_\nu u^\nu}+\gamma \sigma~ r_\nu u^\nu,-\frac{\gamma \alpha {\bm v}}{r_\nu u^\nu}+\gamma \sigma {\bm v}~ r_\nu u^\nu \right),\,
		\label{eq:Av}
		\ea
		where $\gamma=1/\sqrt{1-v^2}$ is the Lorentz factor. Rewriting Eq.~\eqref{eq:Av} in a more compact form, one gets
		\ba
		A^{\mu}=\left(-\frac{\alpha }{(r_\nu u^\nu)}+\sigma~ (r_\nu u^\nu)\right)u^\mu.
		\label{eq:A}
		\ea
		Here, $r_\mu$ is the position four-vector from the heavy quark $(t_r,{\bm x'})$ at retarded time to some field point $(t,{\bm x})$, and $u^\mu=\gamma(1,{\bm v})$ with  $v=|{\bm v}|$. It is important to note that the two events which define $r_\mu$ are connected by a signal propagating at the velocity of the light. Therefore, the events have null separation and $r_\mu$ is a lightlike vector. The modified form of the Cornell potential shown in Eq.~\eqref{eq:A} is similar to the form of the Li\'enard–Wiechert potential in the electrodynamics but the string part is missing there. Now we have,
		\ba
		r_\nu u^\nu=r\gamma-\gamma{\bm r}\cdot {\bm v}=\gamma r\left(1-\hat{\bm r}\cdot {\bm v}
		\right),
		\ea
		where $\hat{\bm r}$ is the unit vector along ${\bm r}$. Then the scalar potential, {\it i.e.,} the zeroth component of the four-potential can be written as
		\ba
		V_{\rm vac}(\bm{r,v})=-\frac{\alpha }{r\left(1-\hat{\bm r}\cdot {\bm v} \right)}+\gamma^2 \sigma\,r\left(1-\hat{\bm r}\!\cdot\! {\bm v} \right).
		\label{eq:Vvac}
		\ea
		This formalism is valid even when the heavy quark velocity is nonuniform. The calculation in this section uses a sequence of independent Lorentz transformations, each performed at a different point along the trajectory of the particle. To fulfill our purpose of heavy quark traveling in QGP medium, we can take the velocity to be constant, and further, if we orient our z-axis along the direction of velocity, then at $t=0$~\cite{Griffiths},
		\ba
		r\left(1-\hat{\bm r}\cdot{\bm v} \right)=\sqrt{z^2+(1-v^2)(x^2+y^2)}.
		\label{eq:ru}
		\ea
		Here, for convenience, the heavy quark is set to pass through the origin at $t=0$ and used the fact that $ r^{\mu}$ is a lightlike vector. Now using Eq.\eqref{eq:ru} in Eq.\eqref{eq:Vvac} the retarded potential in Cartesian coordinates becomes,
		\ba
		V_{\rm vac}(x,y,z,v)&=&-\frac{\alpha }{\sqrt{z^2+(1-v^2)(x^2+y^2)}}\nn 
		&&\hspace{-.8cm}+\,\gamma^2 \sigma~ \sqrt{z^2+(1-v^2)(x^2+y^2)}.
		\ea
		The medium modification of the retarded potential can be done in the Fourier space by dividing the potential with the dielectric permittivity of the medium. Thus, the vacuum potential in Fourier space is obtained as
		\ba
		V_p(p_x,p_y,p_z,v)&=&-\sqrt{\frac{2}{\pi}}\frac{ \alpha}{p_x^2+p_y^2+\left(1-v^2\right) p_z^2}\nn &-&2\sqrt{\frac{2}{\pi}}\frac{\sigma }{ \left[p_x^2+p_y^2+\left(1-v^2\right) p_z^2\right]^2}.
		\ea
		In spherical polar coordinates, the above equation becomes
		\ba
		V_p(\bm{p}, v)&=&-\sqrt{\frac{2}{\pi}}\frac{\alpha}{p^2 \left(1-v^2 \cos ^2\theta \right)}\nn &-&2\sqrt{\frac{2}{\pi}}\frac{ \sigma}{p^4 \left(1-v^2 \cos ^2\theta \right)^2},
		\label{eq:phi}
		\ea
		where $\theta$ is the polar angle in momentum space, i.e., the angle between the $p_z$ and ${\bm p}$. This expression gives the  retarded scalar potential of a moving quark in the vacuum. As discussed earlier, when a charged particle passes through a thermal medium, its properties are affected by the response of that medium. Therefore, when a heavy quark passes through the QGP medium (which is at rest in this scenario), the retarded potential associated with it will be affected by the response of the QGP medium. Therefore, next, we shall discuss the modification of the heavy quark potential given in Eq.~\eqref{eq:phi} through the dielectric permittivity of the QGP medium in the Fourier space. 

		\section{Dielectric permittivity and medium modification of the potential}
		\label{3}
		The medium modified potential in the coordinate space [$ V({\bm r,v})$] can be obtained~\cite{Agotiya:2008ie, Thakur:2012eb} by correcting the vacuum potential with dielectric permittivity encoding the medium screening effect in Fourier space followed by inverse Fourier transformation, i.e.,
		\ba
		V({\bm r,v})=\int \frac{d^3{\bm p}}{(2\pi)^{3/2}} ~(e^{i{\bm p\cdot r}}-1)~\frac{ V_p({\bm p},v)}{\epsilon(p)},\label{vmod}
		\ea
		where $ V_p({\bf p},v)$ is the Fourier transform of the potential in coordinate space 
		and $\epsilon(p)$ is the dielectric permittivity of the medium. Here, we subtract the ${\bm r}-$independent terms in order to renormalize the heavy quark free energy \cite{Dumitru:2009fy}. The inverse of the dielectric permittivity of the static QGP medium is given as ~\cite{Jamal:2017dqs,Beraudo:2007ky},
		\ba
		\epsilon^{-1}(p)=\frac{p^2}{p^2 +m_D^2}-i\pi T \frac{m_D^2 p}{(p^2 +m_D^2)^2},
		\ea
		where $p=|{\bm p}|$ and $m_D$ is the Debye mass of QGP medium obtained from the static limit of longitudinal polarization tensor in the high-temperature limit \cite{Mustafa:2004hf},
		\ba
		{m_D}=T\sqrt{4 \pi  \alpha_s  \left(\frac{N_f}{6}+\frac{N_c}{3}\right)},
		\ea
		where $N_f$ is the number of flavor degrees of freedom and $T$ is the temperature of the medium. We use one-loop strong coupling $\alpha_s$ as~\cite{Laine:2005ai, Srivastava:2015via,Haque:2014rua},  
		\ba 
		\alpha_s =\frac{12\pi}{11N_c -2N_f}\frac{1}{\ln{(\Lambda/\Lambda_{\overline{MS}})^2}},
		\ea
		where $\Lambda_{\overline{MS}}=176 \mbox{MeV}$ is the QCD scale fixing factor and $\Lambda=2\pi T$. 
		
		In order to calculate the exact real part of the potential we decompose the potential into Coulombic part and string part then perform integration separately,
		\ba 
		\Re V({\bm r},v)=\Re V_\alpha({\bm r},v)+\Re V_\sigma({\bm r},v).
		\ea 
		The Coulombic part is written as
		\ba
		\Re V_\alpha({\bm r},v)&=&-\sqrt{\frac{2}{\pi}}\int \frac{d^3{\bm p}}{(2\pi)^{3/2}} \Bigg(\frac{e^{i\boldsymbol{p\cdot r}}}{p^2 +m_D^2}\frac{\alpha}{ 1-v^2 \cos ^2\theta}\nn
		&-&\frac{m_D^2}{p^2 +m_D^2}\frac{\alpha}{p^2 \left(1-v^2 \cos ^2 \theta \right)}\Bigg).
		\ea
		There are no diverging terms in the string part integration, therefore, 
		\ba
		\Re V_\sigma({\bm r},v)&=&-\int \frac{d^3{\bm p}}{(2\pi)^3}(e^{i{\bm p}\cdot {\bm r}}-1)\frac{1}{p^2 +m_D^2}\nn
		&\times&\sqrt{\frac{2}{\pi}}\frac{2 \sigma}{p^2 \left(1-v^2 \cos^2\theta \right)^2}.
		\ea
		In spherical polar coordinates,
		\ba {\bm p} \cdot {\bm r}=r p \big[\sin \Theta  \sin \theta  \cos (\Phi -\phi )+\cos \Theta \cos \theta\big],\nonumber
		\ea
		where the angles $\theta$ and $\phi$ are polar and azimuthal angles in Fourier space (momentum space),  respectively; whereas the angles $\Theta$ and $\Phi$ are polar and azimuthal angles in coordinate space. Since the velocity of the heavy quark is considered along the $z$-axis, $\Theta$ represents the angle between the velocity $\bm{v}$ and the position of the field point $\bm{r}$. The integration over the azimuthal angle, $\phi$ can be done analytically and one obtains,
		\begin{widetext}
			\ba
			\Re V({\bm r},v )&=&-\frac{1}{\pi}\int \frac{\sin\theta\, d\theta\, dp}{p^2+m_D^2} \left[\frac{\alpha\left(m_D^2+p^2e^{i p  r  \cos \theta \cos \Theta } J_0(p\ts r\ts \sin\theta  \sin \Theta )\right) }{ 1-v^2 \cos ^2\theta }\right.\nonumber\\
			&&\hspace{4.5cm}\left.-\frac{2 \sigma  \left(1-e^{i  p  r  \cos \theta \cos \Theta} J_0(p\, r\, \sin \theta \sin \Theta )\right)}{ \left(1-v^2 \cos ^2\theta\right)^2}\right],\label{ReV_ptheta}
			\ea
			where $J_0$ represents the Bessel's function of the first kind. The integration over $p$ and $\theta$ in Eq.~\eqref{ReV_ptheta} can be computed numerically, and the real part of the potential is plotted in Fig.~\ref{NRr}.
			\begin{figure*}[tbh]
				\centering
				\includegraphics[scale=.37]{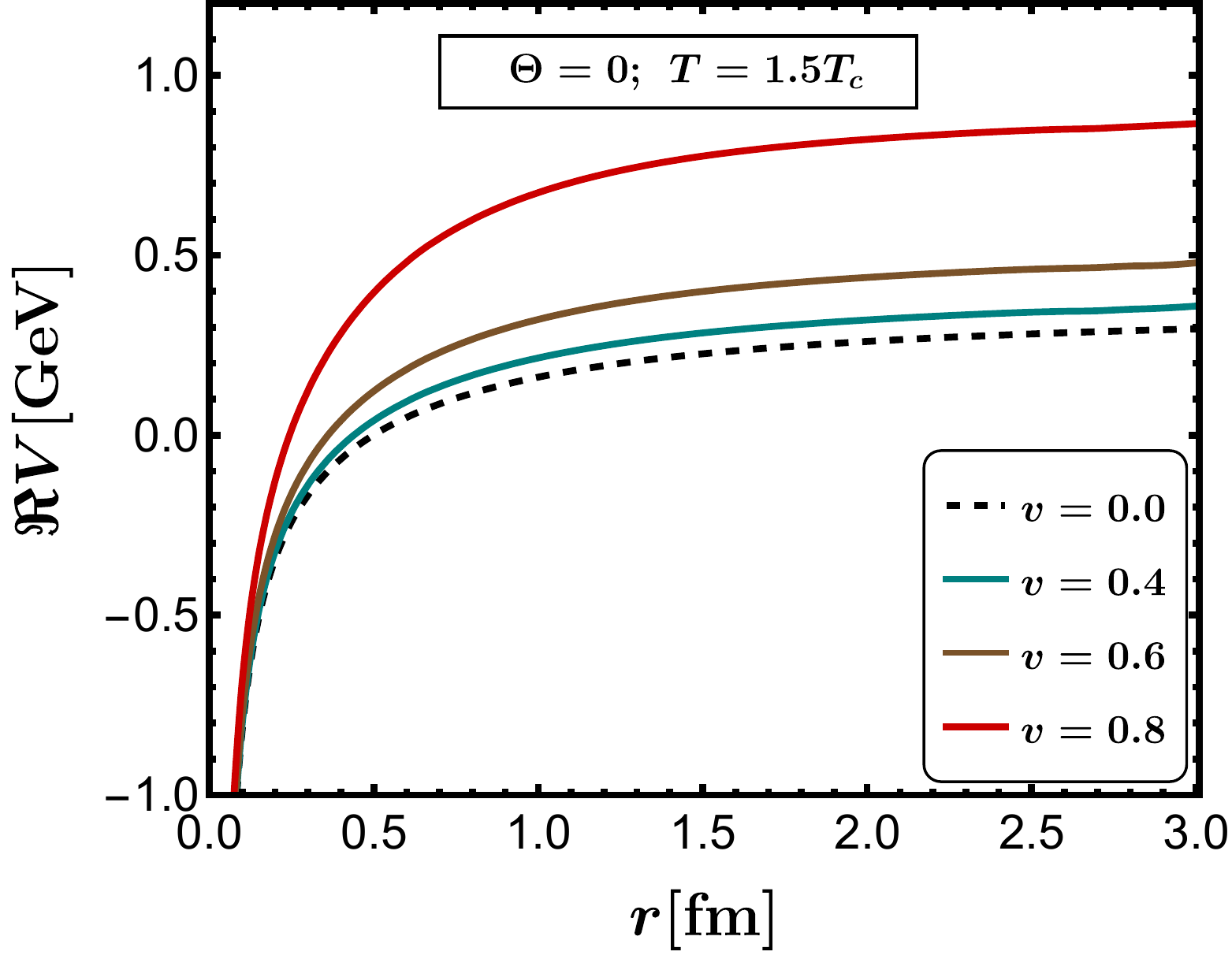}
				\hspace{-2mm}
				\includegraphics[scale=.37]{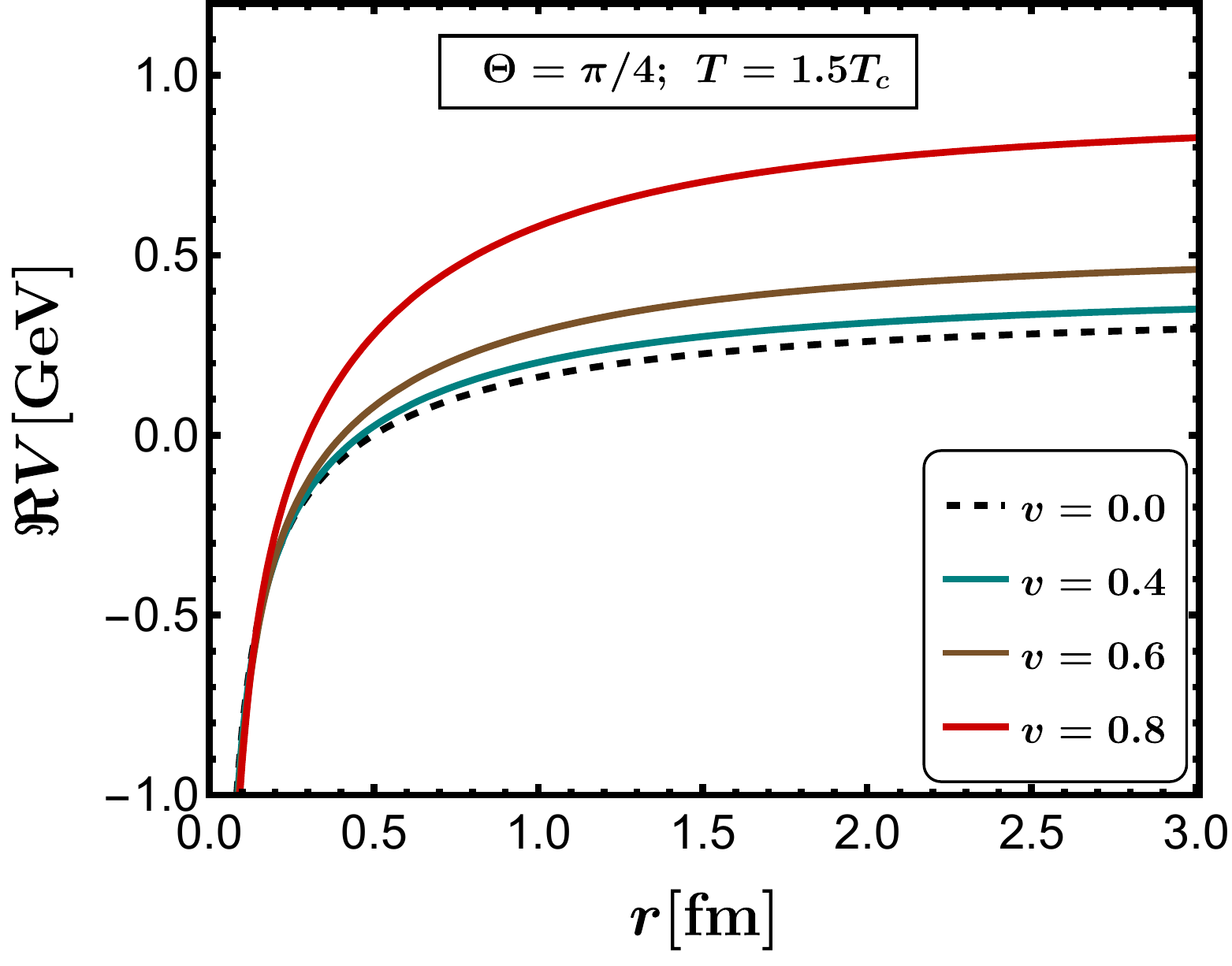}
				\hspace{-2mm}
				\includegraphics[scale=.37]{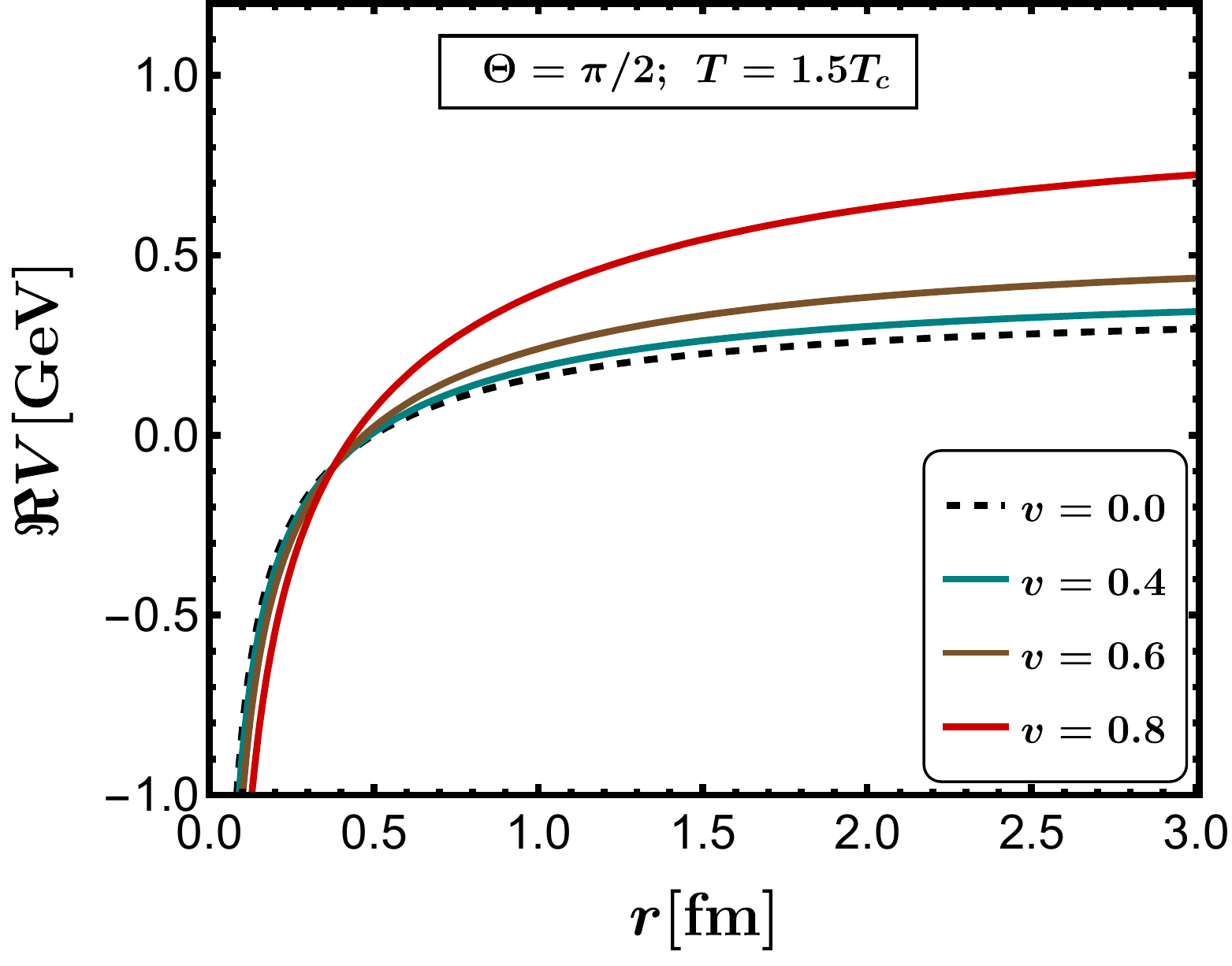}
				\caption{Numerical results for real part of the potential at different velocities and angles [$\Theta =0$ (left ), $\Theta =\pi/4$ (middle ),  $\Theta =\pi/2$ (right)].}
				\label{NRr}
			\end{figure*}
			Similarly, the exact imaginary part of the potential is calculated by substituting the imaginary part of the dielectric function in Eq.~\eqref{vmod},
			\ba
			\Im V({\bm r},v)&=&\int \frac{d^3{\bm p}}{(2\pi)^{3/2}} ~(e^{i\bm{p\cdot r}}-1)~\frac{\pi T m_D^2 p}{\left(p^2 + m_D^2\right)^2 }\sqrt{\frac{2}{\pi}}\left[\frac{\alpha}{p^2 \left(1-v^2 \cos ^2\theta \right)} +\frac{2 \sigma}{p^4 \left(1-v^2 \cos ^2\theta \right)^2}\right].
			\ea
			\begin{figure*}[tbh]
				\centering
				\includegraphics[scale=.37]{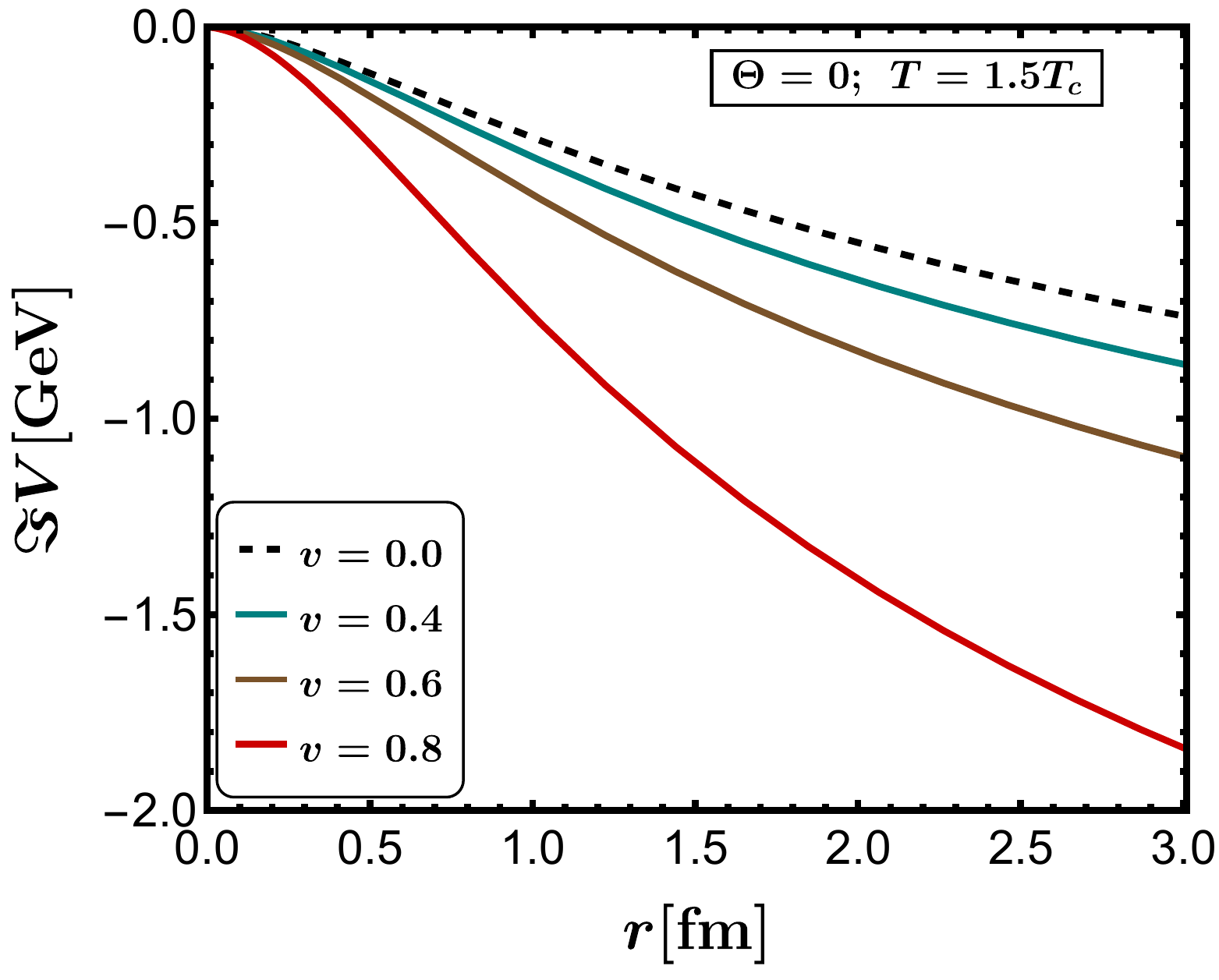}
				\hspace{-2mm}
				\includegraphics[scale=.37]{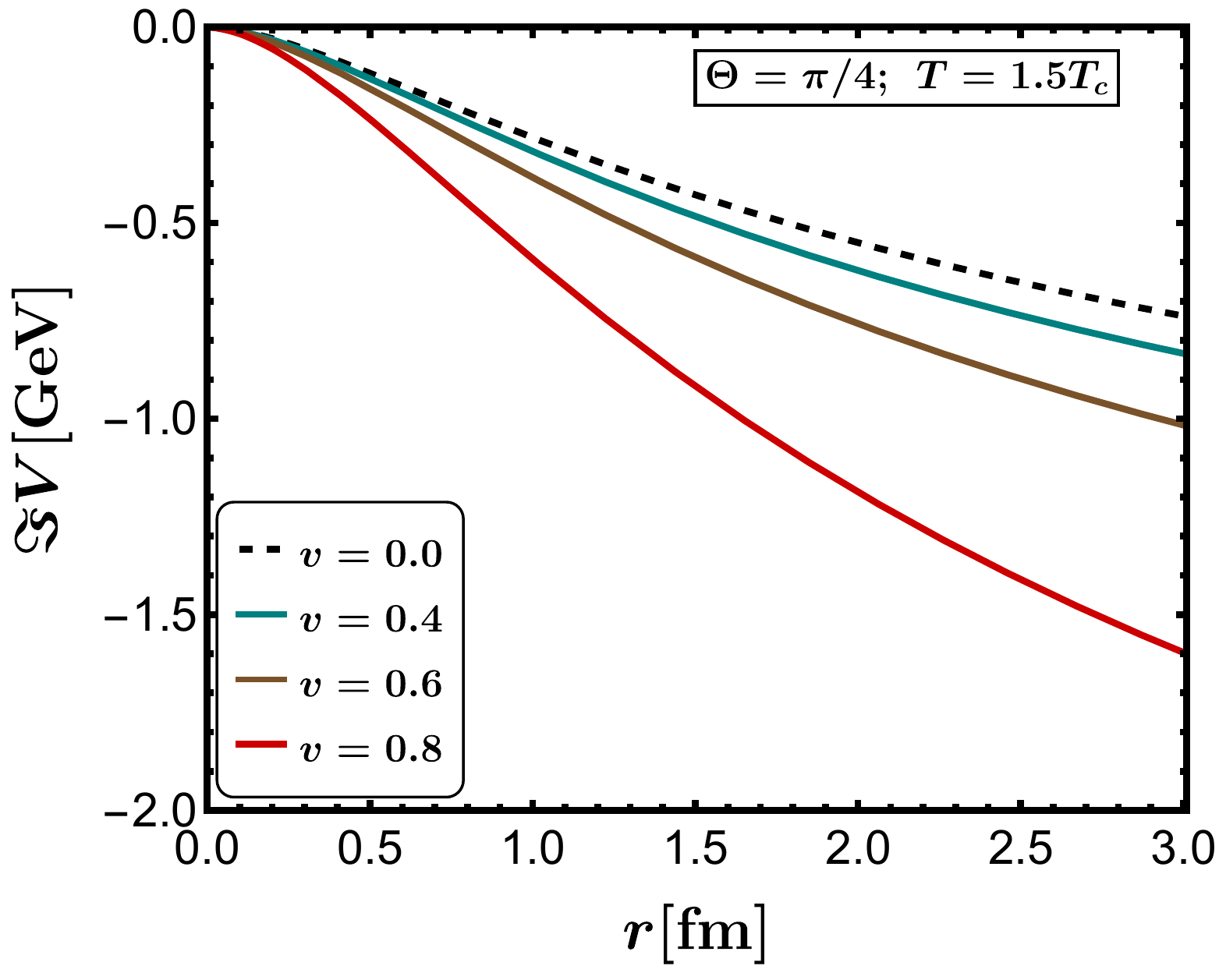}
				\hspace{-2mm}
				\includegraphics[scale=.37]{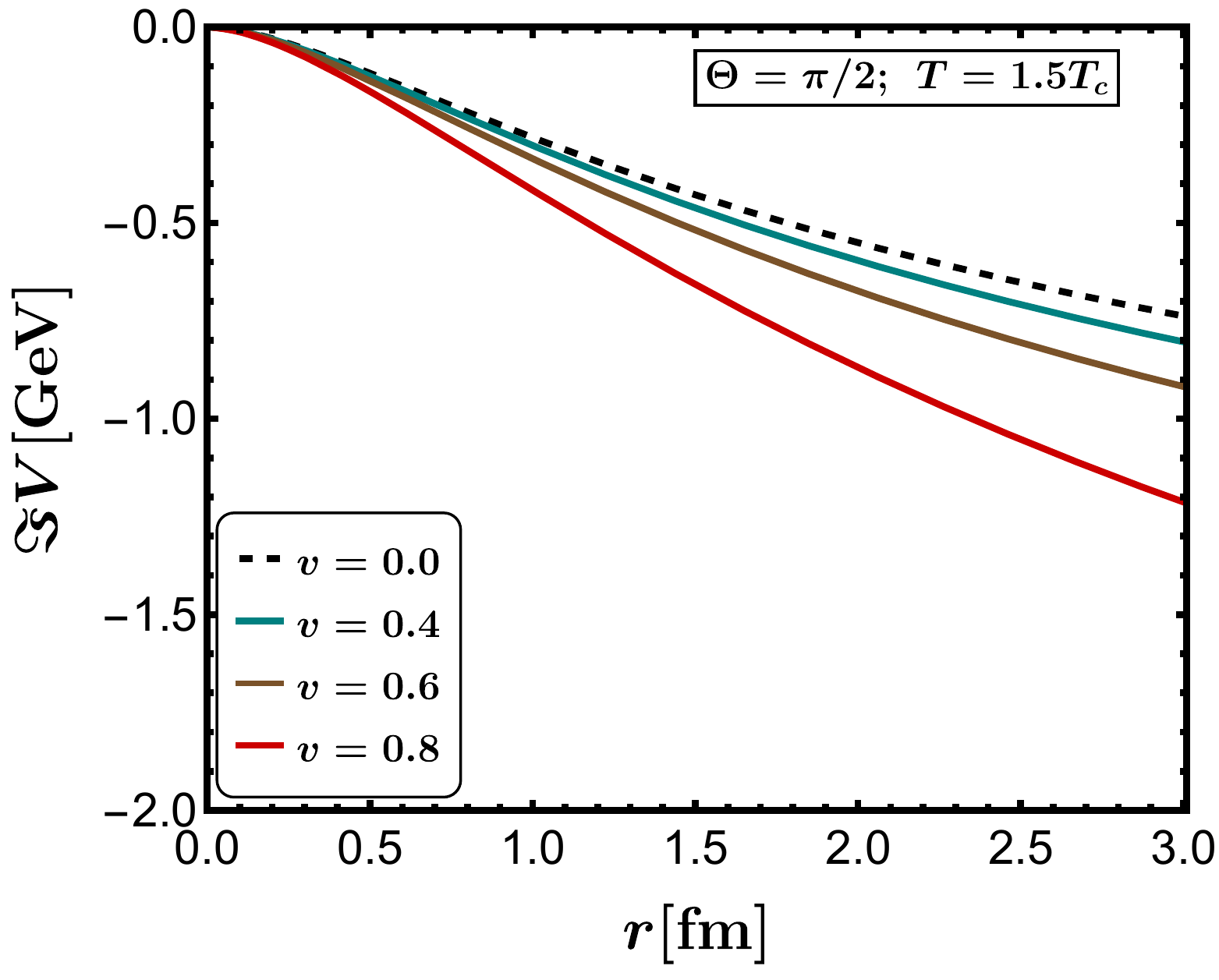}
				\caption{Numerical results for imaginary part of the potential at different velocities and angles [$\Theta =0$ (left ), $\Theta =\pi/4$ (middle ), $\Theta =\pi/2$ (right)].}
				\label{NIr}
			\end{figure*}
			After integrating over $\phi$ we obtain,
			\ba
			\Im V({\bm r},v)&=&-m_D^2T\int\frac{\sin\theta\, d\theta\, dp}{\left(p^2+m_D^2\right)^2}\left[
			\frac{\alpha\ts p }{1-v^2 \cos^2 \theta}+\frac{2\sigma }{ p\left(1-v^2 \cos^2 \theta \right)^2}\right]
			\hspace{0cm}\Big\{1-e^{i p r \cos \theta \cos \Theta} J_0( p r \sin \theta \sin \Theta )\Big\}.\label{ImV_ptheta}
			\ea
			Both the real and imaginary parts of the potentials are independent of $\Phi$ after $\phi$ integration, {\it i.e.,} potential has axial symmetry about the z-axis. The rest of the integration is done numerically, and the results are plotted as shown in the figure for both real (Fig.~\ref{NRr}) and imaginary (Fig.~\ref{NIr}) parts of the potential. 
			\begin{figure*}[tbh]
				\centering
				\includegraphics[scale=0.46]{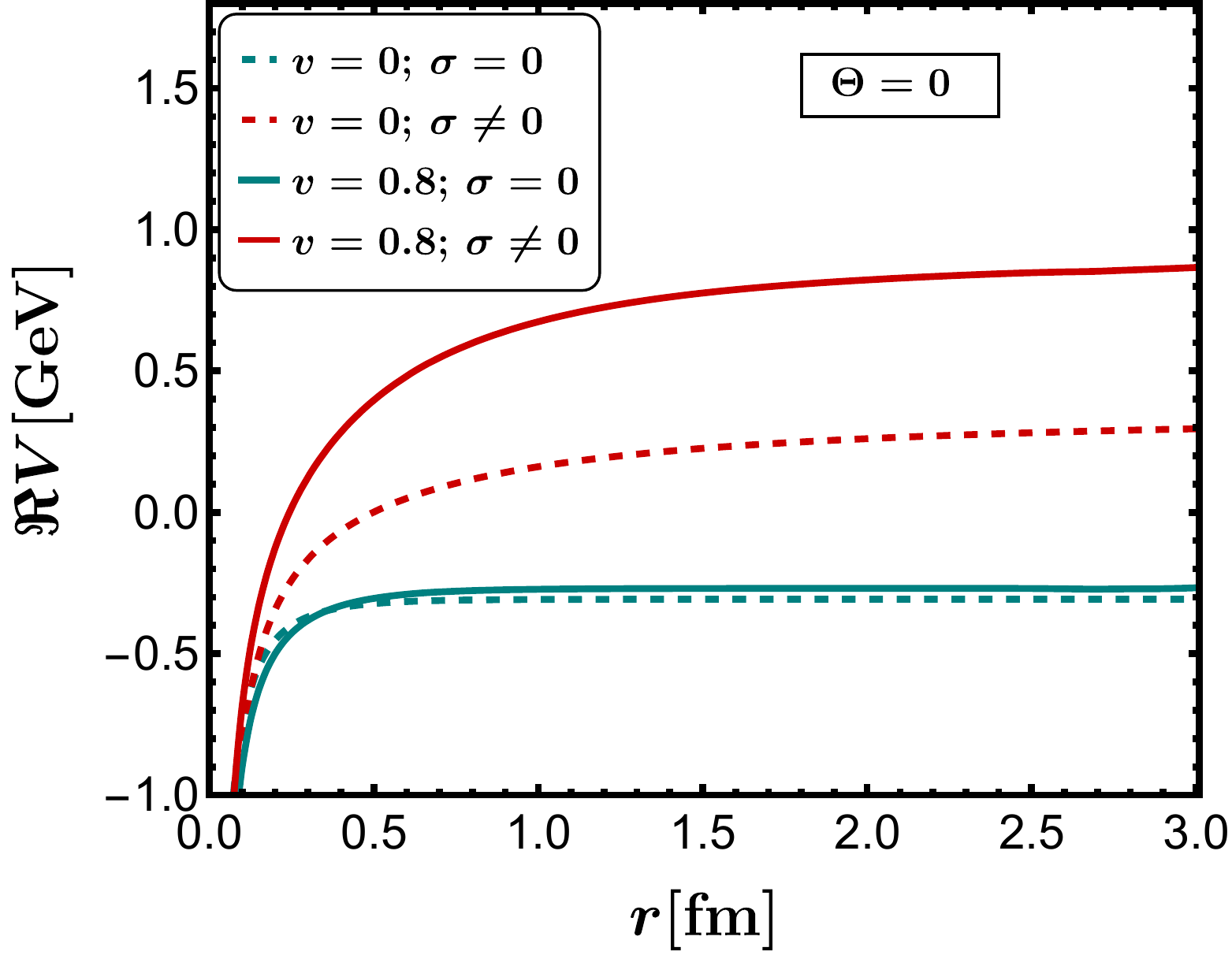}
				\hspace{10mm}
				\includegraphics[scale=0.46]{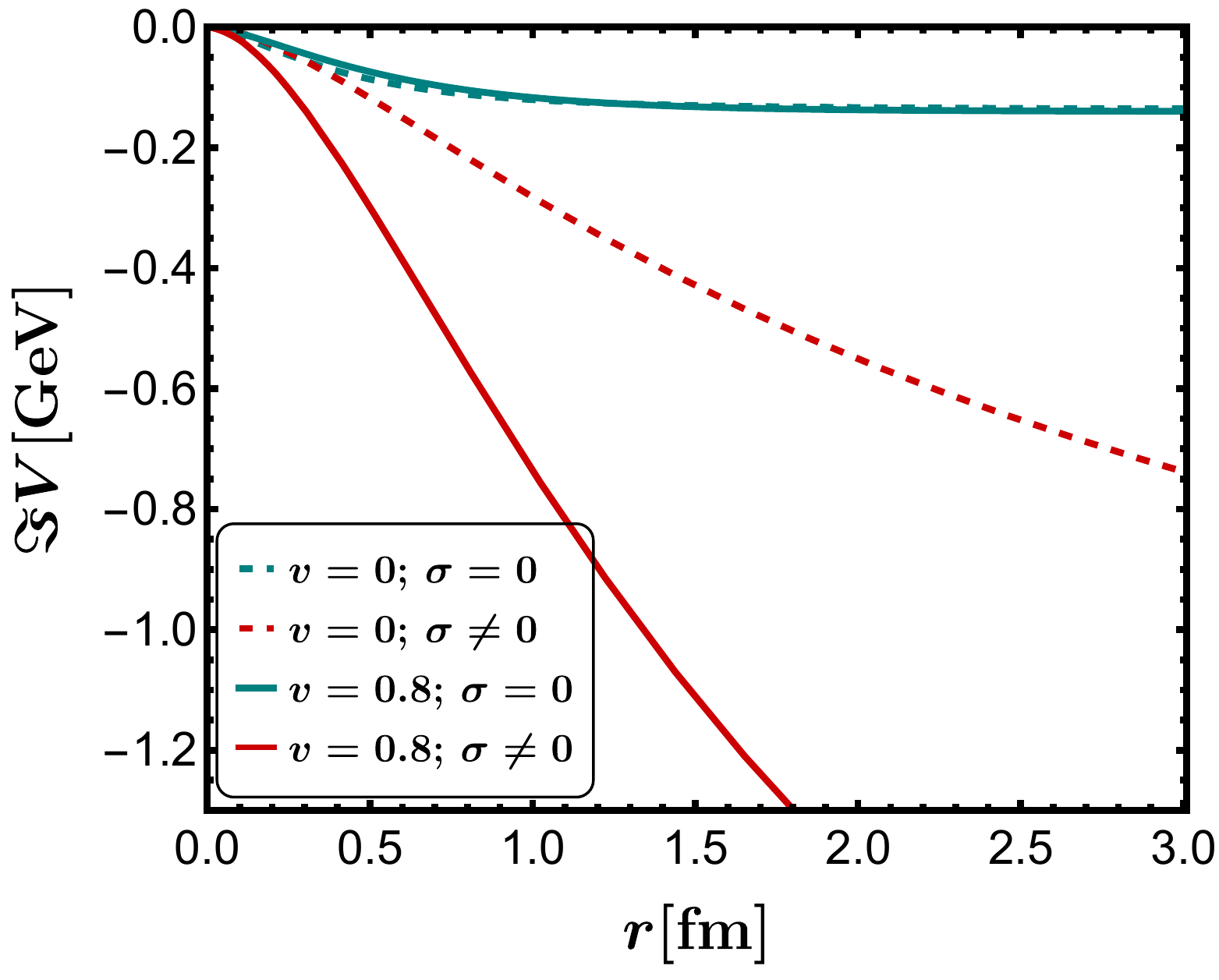}
				\caption{Numerical results for real (left) and imaginary (right) parts of the potential, a comparison of Cornell and Coulomb potential. }
				\label{Sigma_comp}
			\end{figure*}
			\begin{figure*}[tbh]
				\centering
				\includegraphics[scale=0.46]{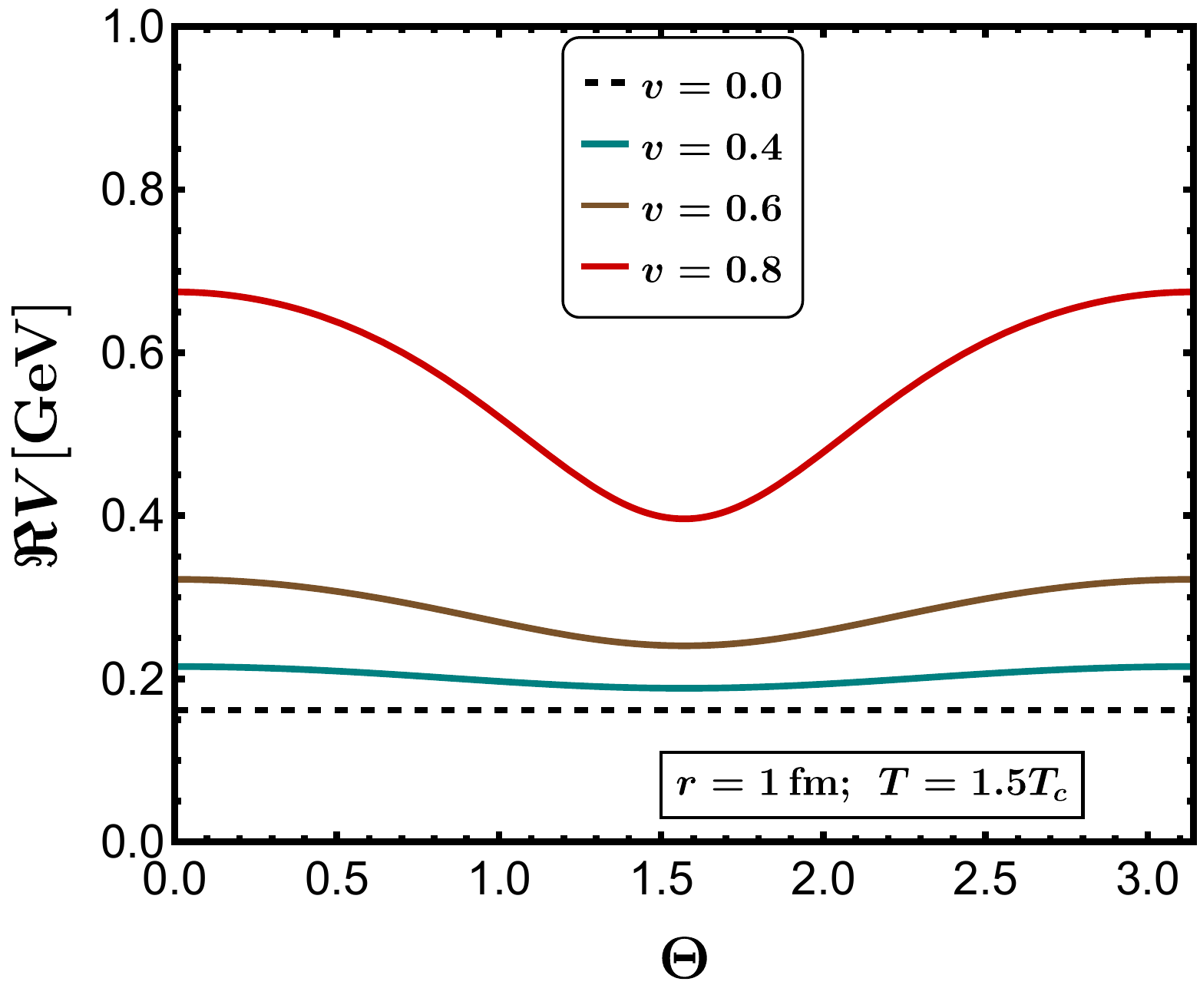}
				\hspace{10mm}
				\includegraphics[scale=0.46]{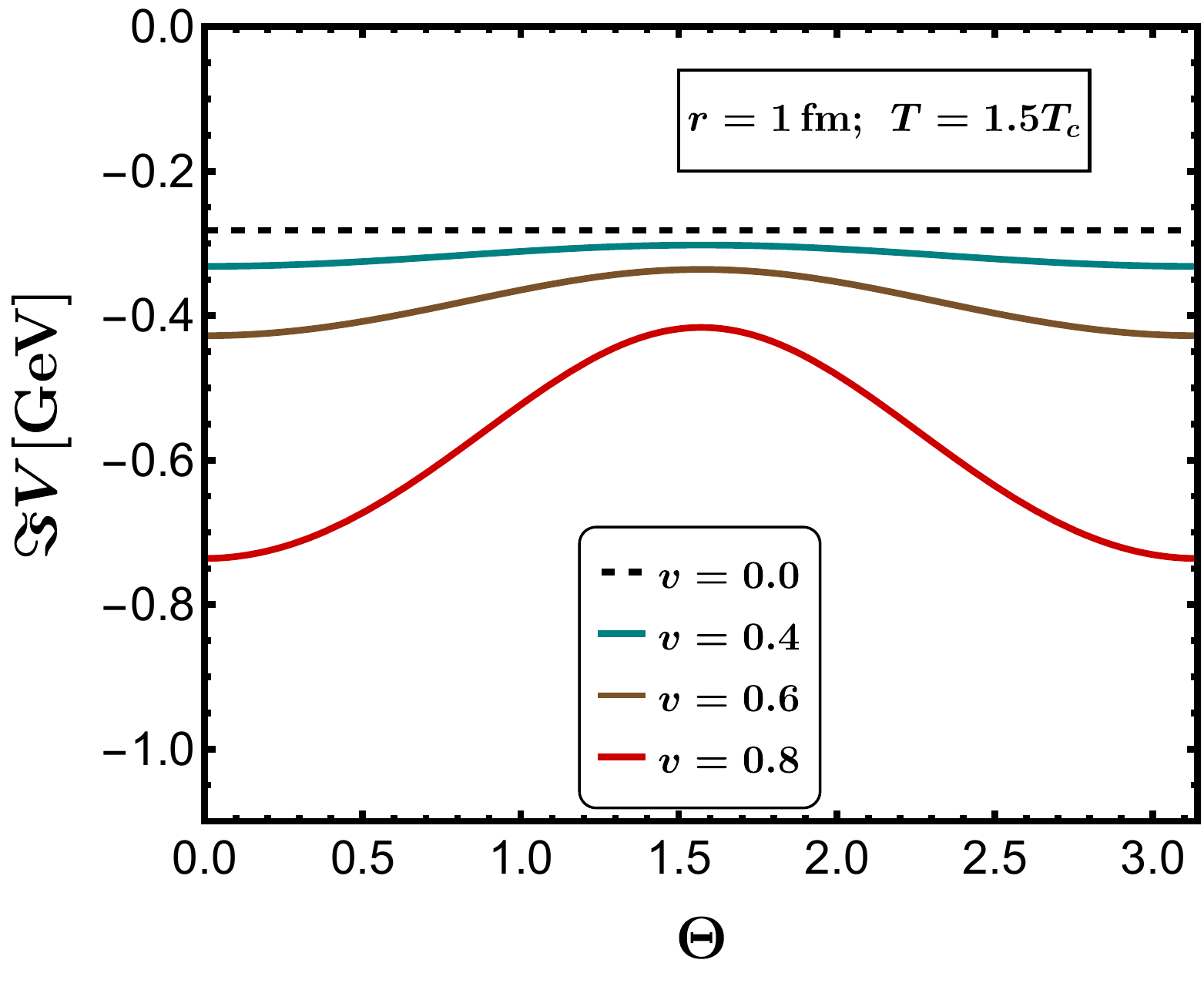}
				\caption{Numerical results for the angular variation of real (left) and imaginary (right) parts of the potential at different velocities. Here, $\Theta$ is in radians }
				\label{Nangular}
			\end{figure*}

			\begin{figure*}[tbh]
				\centering
				\includegraphics[scale=0.46]{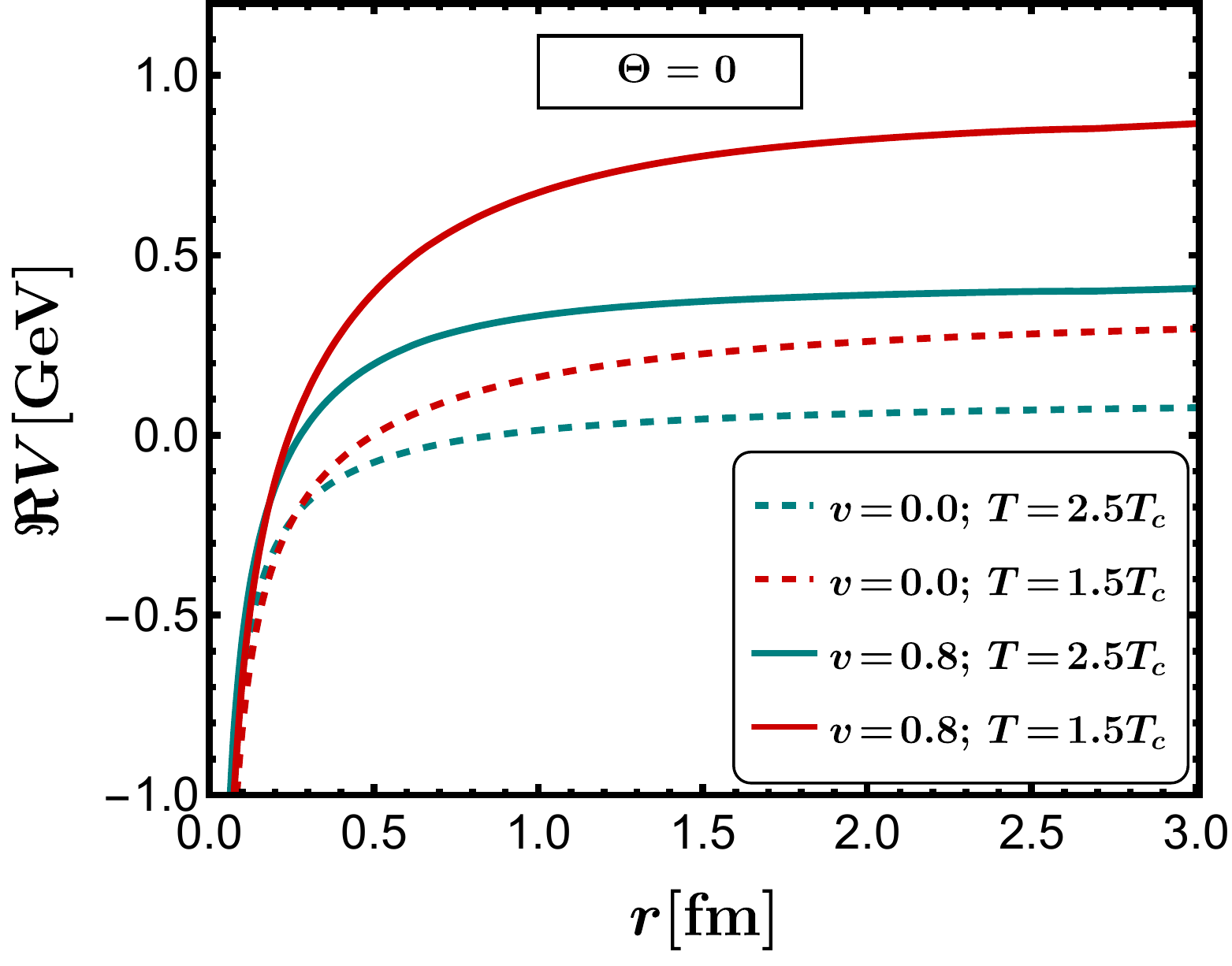}
				\hspace{7mm}
				\includegraphics[scale=0.46]{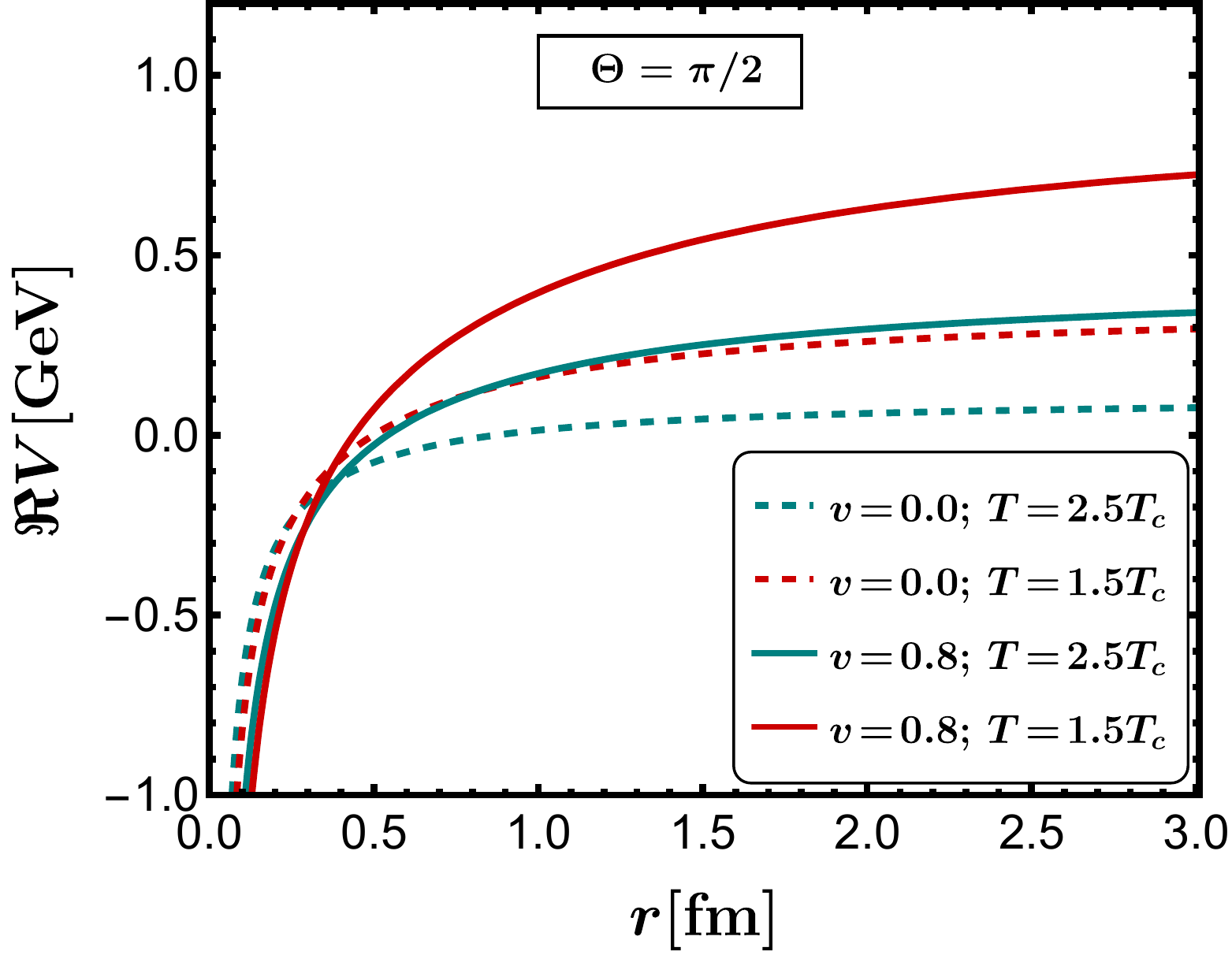}
				\caption{Numerical results for real potential at different temperatures and velocities, a comparison. }
				\label{TVRcomp}
			\end{figure*}
			\begin{figure*}[tbh]
				\centering
				\includegraphics[scale=0.46]{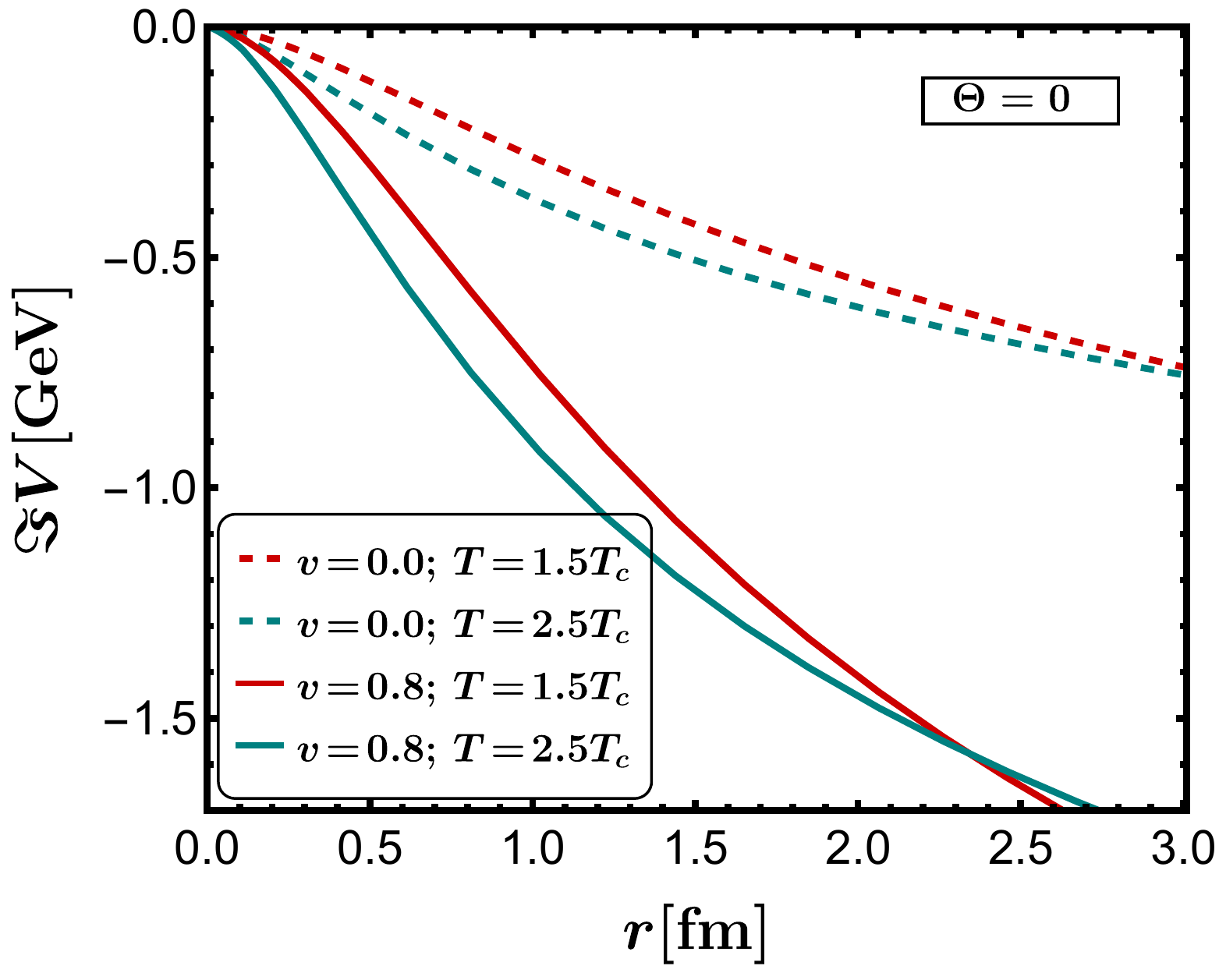}
				\hspace{7mm}
				\includegraphics[scale=0.46]{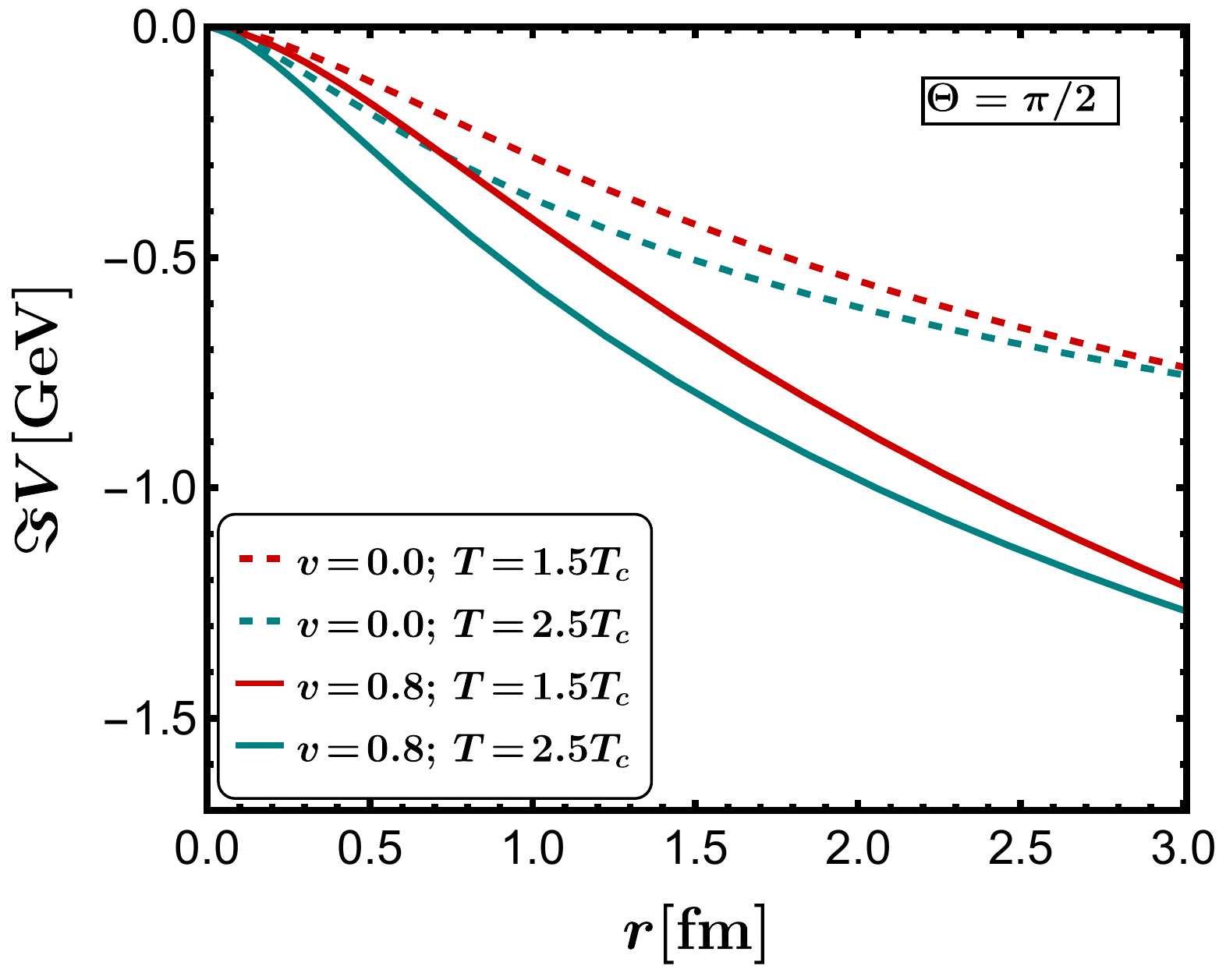}
				\caption{Numerical results for imaginary potential at different temperatures and velocities, a comparison. }
				\label{TVIcomp}
			\end{figure*}
		\end{widetext}
		\section{Potential at small velocities}
		\label{4}
		The real and imaginary parts of the potential as obtained in Eqs.~\eqref{ReV_ptheta} and~\eqref{ImV_ptheta} can be simplified at small velocities. Considering small velocity, one can  expand the $ V_p(\bm{p}, v)$ in Eq.~\eqref{eq:phi} and keep terms up to $O\left(v^2\right)$ as,
		\ba
		\frac{1}{1-v ^2 \cos ^2 \theta}\approx 1+v^2 \cos ^2 \theta + O\left(v^4\right).\label{approx}
		\ea
		This approximation is valid for the case of a  quarkonia moving in the QGP medium with relatively small velocity. Next, we can analytically perform the integration in Eq.~\eqref{vmod} using the approximation given in Eq.\eqref{approx} to obtain the real part and imaginary part of the potential. 
		The modified form of the potential in Eq.~\eqref{eq:phi} in small velocity limit is
			\begin{widetext}
		\ba
		V_p(p,\theta,v)&=&-\sqrt{\frac{2}{\pi}}\frac{1}{p^2}\Big(\alpha +\alpha   v^2 \cos ^2 \theta\
		\left.\hspace{0cm}+\frac{2 \sigma }{p^2}+\frac{4 \sigma  v^2  \cos ^2\theta}{p^2}\right).\label{smallv}
		\ea
		Therefore the real part of the potential at small velocity is obtained as
		\ba \Re V(\bm{r}, v)=\int \frac{d^3{\bm p}}{(2\pi)^{3/2}} ~(e^{i{\bm p\cdot r}}-1) V_p(p,\theta,v) \rm{Re}[\epsilon^{-1}(p)].\nn\label{approxreal}
		\ea 
		The integration in Eq.~\eqref{approxreal} is easy to calculate in spherical polar coordinates with $\cos \theta=p_z/p$. Doing so, the real part of the potential is computed as,
			\ba
			\Re V(\rho, \Theta, v)\hspace{-.0cm}&\approx&\hspace{-.0cm}-\frac{\alpha m_De^{-\rho}}{\rho}-\alpha m_D - \frac{2  \sigma}{m_D \rho}\left(1- e^{- \rho}\right)+\frac{2\sigma }{m_D}
			-\frac{\alpha v^2 m_D}{3}-\frac{\alpha  m_D v^2}{ \rho^3}\left(1-3 \cos^2 \Theta \right)\nn
			&&
			+\ \frac{4 \sigma v^2}{3 m_D}+\frac{\alpha  m_Dv^2  e^{-\rho}}{\rho}\left\{\frac{1}{ \rho}+\frac{1}{\rho^2}\right.
			\left.-\left(1+\frac{3 }{\rho}+\frac{3}{ \rho^2}\right)\cos^2 \Theta\right\} -\frac{2  \sigma  v^2 }{m_D \rho}\left(1- \cos^2 \Theta \right)\
			\nn
			&&+\ \frac{4  \sigma  v^2}{ m_D \rho^3}\bigg(1-3 \cos^2 \Theta \bigg)+\frac{4  \sigma  v^2  e^{-\rho}}{m_D\rho}\bigg(\cos^2 \Theta -\frac{1-3 \cos^2 \Theta }{\rho}-\frac{1-3 \cos^2 \Theta }{ \rho^2}\bigg),\hspace{.7cm} \label{ReV_approx}
			\ea
		\end{widetext}
		where $\rho=m_Dr$. At $v=0$, the approximate real part of the potential in Eq.~\eqref{ReV_approx} becomes the more familiar screened Cornell potential where the angular dependence is also disappeared,
		\ba
		\left.\Re V(r)\right|_{v=0}&=&-\frac{\alpha e^{-{m_D} r}}{r}-\alpha m_D\nn
		&&-\frac{2  \sigma}{m_D^2 r}\left( 1- e^{-{m_D} r}\right)+\frac{2\sigma }{m_D}.\label{ReV_v=0}
		\ea
		The screened Cornell potential at $v=0$ further converges to the vacuum Cornell potential as $m_D\rightarrow 0$ with  $T\rightarrow 0$, we have
		\ba
		\left.\Re V(r)\right|_{v=0; T=0}&=&-\frac{\alpha }{r}+\sigma r.
		\ea
		Similarly, the imaginary part of the retarded potential in the small velocity approximation can be obtained as,
		\ba
		\Im V(\bm{r},v)=\int \frac{d^3{\bm p}}{(2\pi)^{3/2}} ~(e^{i{\bm p\cdot r}}-1) V_p(p,\theta,v) \rm{Im}[\epsilon^{-1}(p)].\nonumber\\
		\label{approximag}
		\ea
		After performing the $\phi$ and the $\theta$ integration, we obtain the following results for the imaginary part of the potential. For the static case, the potential is isotropic, and it is obtained as 
		\ba
		&&\hspace{-1cm}\Im V_{\rm iso}=-2 \alpha  T  \int\limits_{0}^{\infty}\frac{z}{\left(z^2+1\right)^2} \left[1-\frac{\sin ({m_{D} r} z)}{{m_{D} r} z}\right]dz \nn
		&-& \frac{4 \sigma T}{m_{D}^2} \int_{0}^{\infty}\frac{dz}{z \left(z^2+1\right)^2}\left[1-\frac{\sin ({m_{D} r} z)}{{m_{D} r} z}\right],\label{eq:iso}
		\ea
		where $z=p/m_D$. In the small velocity limit, the imaginary part of the potential can be expressed as~\cite{Thakur:2012eb, Dumitru:2009ni}
		\ba
		\Im V = \mathcal{A}(r,T,v) + \mathcal{B}(r,T,v) \cos(2\Theta).\label{eq_IVAB}
		\ea
		In general, one can proceed with any angle $\Theta$ and evaluate the integration over $p$ numerically. Here we are showing the results for $\Theta=0$ (parallel case) and $\Theta=\pi/2$ (perpendicular case) in small velocity limits as 
		\begin{widetext}
			\ba
			\Im V^\parallel\left(v,r\right)= \Im V_{\rm iso}+ \frac{2}{3} v^2 T\Bigg(\frac{\alpha }{3} \int_{0}^{\infty}\frac{zdz}{\left(z^2+1\right)^2}+  \frac{4\sigma}{{m_D}^2} \int_{0}^{\infty}\frac{dz}{z \left(z^2+1\right)^2}\Bigg) \left[1-\frac{3 \sin ( \rho z)}{{\rho} z}-\frac{6\cos ( \rho z)}{ \rho^2 z^2}+\frac{6 \sin (\rho z)}{ \rho^3 z^3}\right],
			\label{eq:VI0}
			\ea
			and
			\ba
			\Im V^\perp\left(v,r \right)= \Im V_{\rm iso}+ \frac{2}{3} v^2 T\Bigg(\frac{\alpha }{3} \int_{0}^{\infty}\frac{z dz}{\left(z^2+1\right)^2} + \frac{4\sigma}{{m_D}^2} \int_{0}^{\infty}\frac{dz}{z \left(z^2+1\right)^2}\Bigg)\times\left[1+\frac{3 \cos (\rho z)}{ \rho^2 z^2}-\frac{3 \sin (\rho z)}{ \rho^3 z^3}\right].
			\label{eq:VI90}
			\ea
		\end{widetext}
		Therefore we can write
		\ba
		\mathcal{A}(r,T,v)=\left[\Im V^\parallel\left(v,r \right)+\Im V^\perp\left(v,r \right)\right]/2
		\ea and
		\ba\mathcal{B}(r,T,v)=\left[\Im V^\parallel\left(v,r \right) - \Im V^\perp\left(v,r \right)\right]/2.
		\ea
		It is evident from Eq.~\eqref{eq:VI0} and Eq.~\eqref{eq:VI90} that at $v=0$, the approximate imaginary part of the potential will contain only the isotropic part given in Eq.~\eqref{eq:iso}, which also vanishes in the vacuum as $T\rightarrow 0$. That means only the Cornell potential given in Eq.~\eqref{CORNELLREST} remains after taking the limit $v\rightarrow 0$ and $T\rightarrow 0$, the original potential we started with.

		\section{Thermal Width}
		\label{5}
		The thermal width of the quarkonium resonant state can be studied from the imaginary part of the potential. We will consider that the imaginary part of the potential is the perturbation to the vacuum potential, and we calculate the thermal width in the first order of perturbation as~\cite{Song:2007gm, BitaghsirFadafan:2015yng,Thakur:2016cki}
		\ba
		\Gamma_{Q\bar{Q}}(v)=-\langle\Psi|\Im V\left(v,r,\Theta \right)|\Psi\rangle,\label{width}
		\ea
		where $\Psi(r)$ is the wave function of the quarkonium bound states. The leading contribution to the potential for the deeply bound quarkonium states in QGP is Coulombic. Therefore, the hydrogen atom wave function is a good approximation in this context to calculate the thermal width of quarkonium bound states. Now, quarkonium wave function in the QGP frame is
		\ba
		\Psi(r)=\frac{1}{\sqrt{\pi a_0^3}}e^{-q/a_0},\label{wavefunction}
		\ea
		 where $q=r\sqrt{1+\frac{v^2 \cos ^2\Theta }{1-v^2}}$ is due to the Lorentz transformation of the wave function, $a_0=2/(C_F m_Q \alpha_s)$ is the Bohr radius corresponds to the quarkonia and $m_Q$ is the quark mass. Note that one can get the exact wave function solving Schr\"odinger equation with the real part of the potential~\eqref{ReV_ptheta} and we intended to do that in near future. Substituting Eq.~\eqref{wavefunction} in Eq.~\eqref{width} gives 
		\ba
		\Gamma_{Q\bar{Q}}(v)= -\frac{1}{\pi a_0^3}\int d^3 r e^{-2q/a_0}\Im V\left(r,v,\Theta\right).
		\label{eq:tw}
		\ea
		Here we obtain the exact results using full imaginary potential given in Eq.~\eqref{ImV_ptheta} as
		\begin{widetext}
			\ba
			\Gamma_{Q\bar{Q}}(v)&\!\!=\!\!&\frac{2 m_D^2T}{a_0^3}\int dr \, d\Theta \, r^2 \sin\Theta e^{-2q/a_0} \int \frac{\sin\theta\, d\theta\, dp}{\left(p^2+m_D^2\right)^2}
			\times\left[
			\frac{\alpha\ts p }{1-v^2 \cos^2 \theta}+\frac{2 \sigma }{ p\left(1-v^2 \cos^2 \theta \right)^2}\right].\hspace{.6cm}
			\label{wid_vT}
			\ea
		\begin{figure}[tbh]
			\centering
			\includegraphics[scale=0.47]{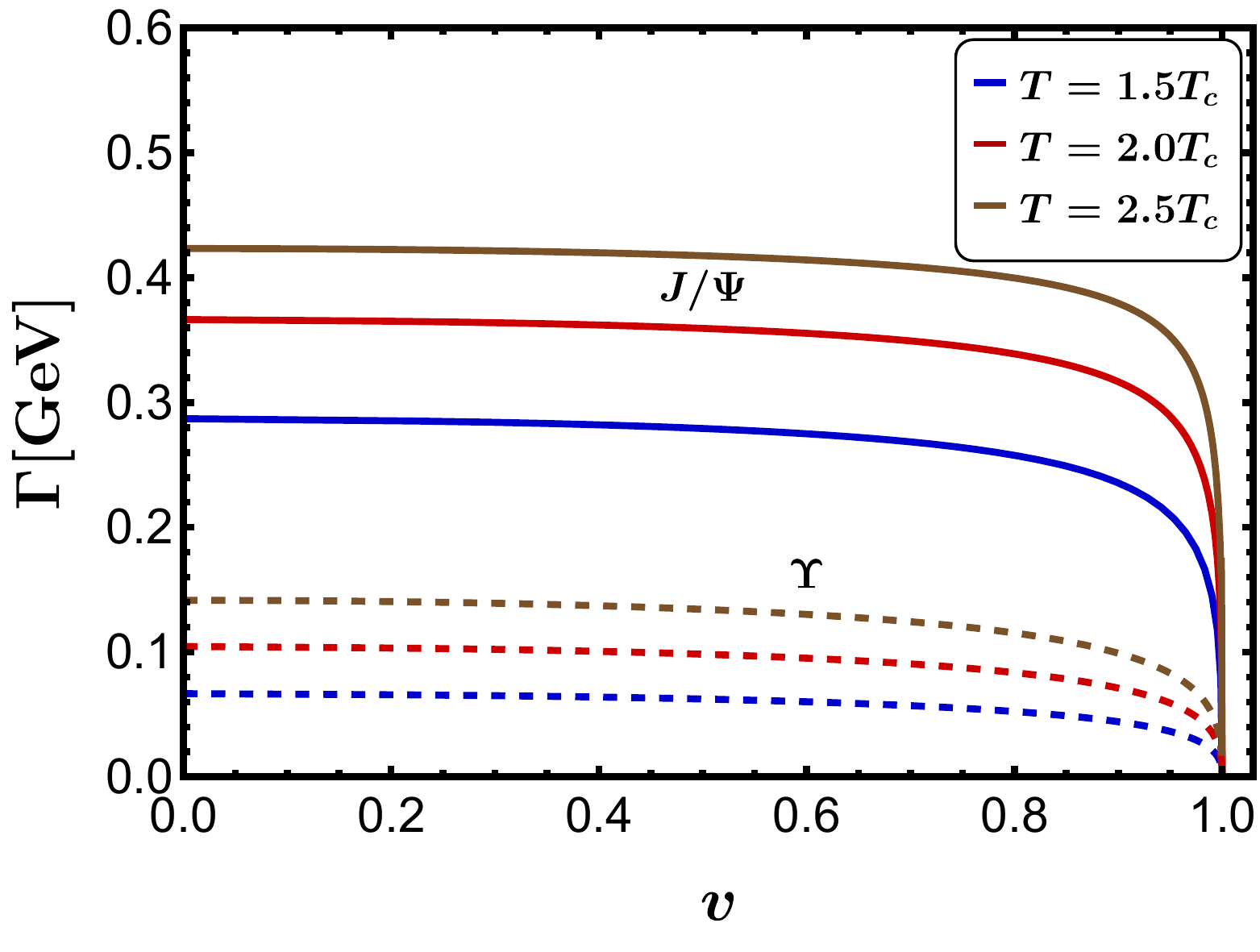}
			\hspace{2mm}
			\includegraphics[scale=.47]{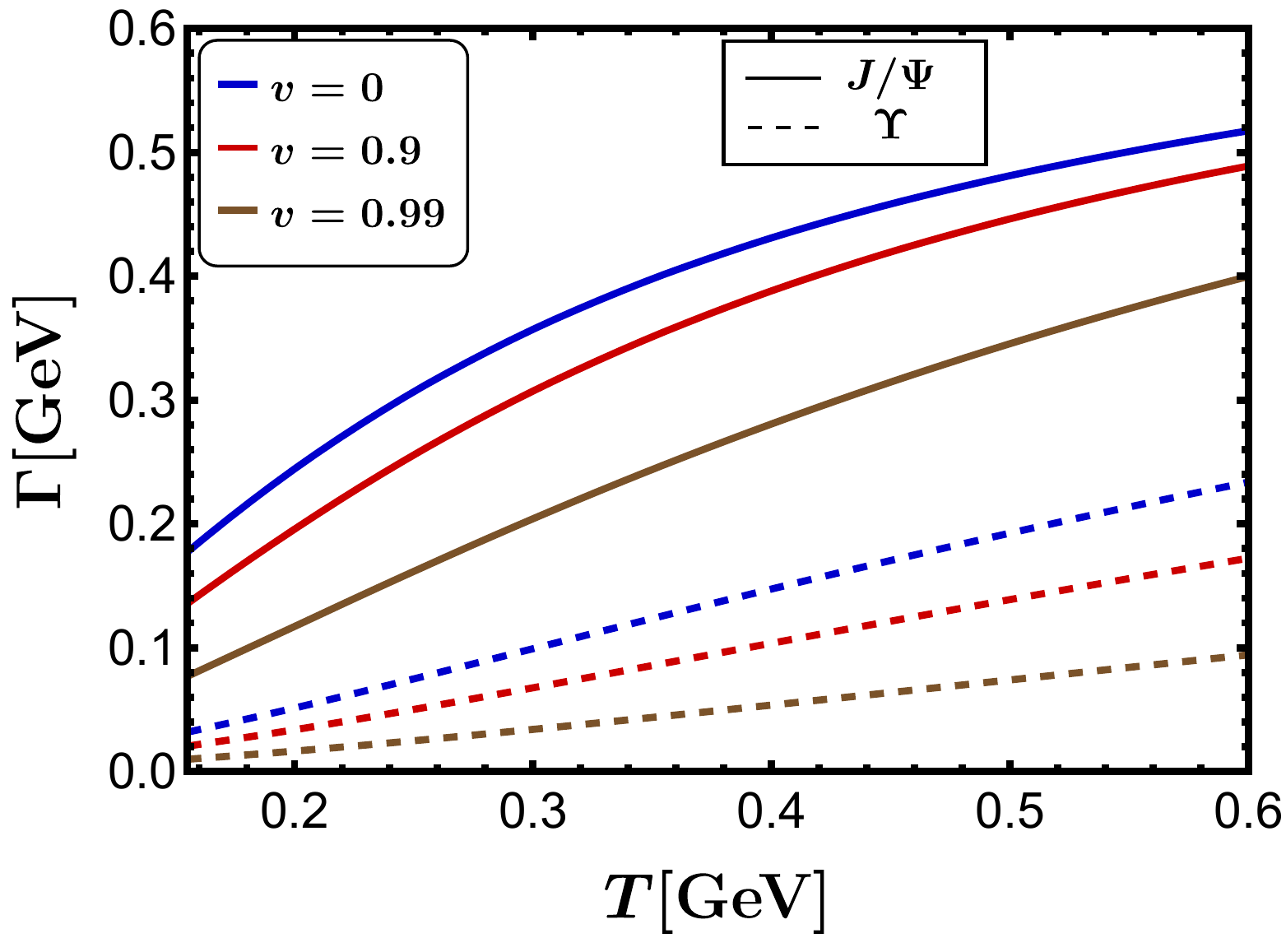}
			\caption{Decay width of the $J/\psi(1s)$  and $\Upsilon(1s)$  with velocity (left) and temperature (right).}
			\label{widthvel}
		\end{figure}	
		\end{widetext}
		We plot the thermal width of the charmonium and the bottomonium ground states as a function of temperature and velocity using the expression derived above, which are further discussed in the result sections. 
		\section{Results}
		\label{6}
		The heavy-quarkonium potential in the QGP medium is studied analytically and numerically with respect to various parameters, primarily the quarkonia velocity and angular dependence. The thermal width of the quarkonium ground state is obtained, and its dependence on velocity and temperature is studied. The illustration of results in various plots used different temperature $T=1.5\,T_c,~2\,T_c$, and  $2.5\,T_c$ where the crossover temperature, $T_c=0.155~\rm{GeV}$. The number of quark flavors $N_f = 3$ and $\sigma=0.18~\mbox{GeV}^2$. The temperature dependence in the potential arises through the strong coupling ($\alpha_s$), dielectric function $\epsilon(p)$, and Debye mass ($m_D$).
		
		Figure~\ref{NRr} shows the variation of the real part of the potential with distance $r$ at angle $\Theta=0$ (left), $\Theta=\pi/4$ (middle), and $\Theta=\pi/2$ (right) and temperature $T=1.5~T_c$. We can observe that the potential and its variation are different in all three directions. Initially, the potential increases sharply and then saturate as the distance increases. The potential decreases with an increase in velocity at a very short distance, whereas at a large distance, the potential increases as velocity increases; this switching is more noticeable in the $\Theta=\pi/2$ case in Fig.~\ref{NRr}. At small velocities, the deviation of the potential from the static case ($v=0$) is very small, but as the velocity becomes very high, a rapid shift in potential is observed. As distance $r$ increases, the potential becomes positive, and this sign flipping happens quickly as velocity increases. This means the negative potential region is less for fast-moving heavy quarks, and the probability of quarkonia formation is less. Also, as we move from $\Theta=0$ to $\Theta=\pi/2$, the potential becomes positive rather slowly.
		Similarly, Fig.~\ref{NIr} shows the imaginary part of the potential against $r$ at the same parameters mentioned above. Here, the imaginary potential is always negative as expected, and its magnitude increase as velocity increases. The quarkonium potential is more sensitive to the velocity along the direction of motion, i.e., at $\Theta=0$.  
		
		Figure~\ref{Sigma_comp} illustrates the comparison between Cornell potential and Coulomb potential ($\sigma=0$ case) along the direction of motion of the heavy quark. Both the real and imaginary parts of the Coulomb potential have nominal dependence on velocity, whereas the string part of the potential have substantial  dependence on velocity from static to relativistic case. This implies the velocity dependence of the Cornell potential is almost solely due to the string part of the potential.
		
		Figure~\ref{Nangular} shows the angular dependence of the real (left) and imaginary (right) part of the potential at $r\!=\!1\, \rm{fm}$ and $T=1.5~T_c$, respectively. Both parts are symmetric about the plane containing the particle and perpendicular to the direction of motion. It is interesting to note that velocity dependence is most prominent along the direction of the velocity of heavy quark for both real and imaginary parts of the potential. At low velocities, there is little variation in potential, but as velocity increases, the spherical symmetry breaks down, leading to an increase in anisotropy. The real part is minimum, and the imaginary part is maximum at $\Theta=\pi/2$ direction. Therefore, the quarkonia are most likely to be oriented in a plane perpendicular to its direction of motion.
		
		In Fig.~\ref{TVRcomp} we have made a comparison between velocity dependence and temperature dependence of the real part of the potential. It is interesting to note that the effect of velocity decreases with an increase in temperature in the case of the real part of the potential, especially at $\Theta=0$, i.e., the variation of potential with change in velocity is more at $T=1.5~T_c$ than $T=2.5~T_c$. In comparison to the static case, $v=0$, the potential changes more at finite/high velocity with temperature. Our results show that the velocity dependence of the real part is as important as temperature dependence.
		Similarly, in Fig.~\ref{TVIcomp} we compare velocity dependence and temperature dependence of the imaginary part of the potential. The variation of the potential with temperature at different angles and different velocities is more or less the same. The potential changes rapidly along the direction of motion $\Theta=0$ of the heavy quark than the perpendicular direction $\Theta =\pi/2$. Our results show that the heavy quark velocity, as well as the medium temperature, highly influence the quarkonium potential.
		
		Figure~\ref{widthvel} shows the variation of the thermal width with velocity (left) and temperature (right) of charmonium $(J/\Psi)$ and bottomonium $(\Upsilon)$ ground states. Even though the magnitude of the imaginary part of the potential increases with velocity, the thermal width decrease with velocity due to the phenomenon of time dialation. Note that the thermal width obtained here qualitatively agrees with the thermal width calculated in  Ref.~\cite{Thakur:2016cki} within the real-time formalism using the hard thermal loop approximation and also with results in Ref.~\cite{Song:2007gm} at the leading order in perturbative QCD. The width increases with temperature for the both charmonium and bottomonium states. The mass of the charm quark (taken $M_{c}= 1.27 ~\rm{GeV}$) is less as compared to the bottom (taken $M_{b}= 4.18 ~\rm{GeV}$), one can notice that the thermal width of $J/\Psi$ is higher than $\Upsilon$ for the same parameters. This preserves the fact that the lighter bound state, i.e., $c{\bar c}$ dissociates faster than the comparatively heavier one.
		
		\section{Summary and conclusion}
		\label{7}
		In the current analysis, we have studied the potential of a moving heavy quarkonium in a static QGP medium. First, we derived the retarded potential of a uniformly moving heavy quark in the vacuum following the analogy of the Li\'enard-Wiechert potential in the electrodynamics, where we performed Lorentz transformation on the static potential to find its form in a boosted frame. The resulting velocity and angular-dependent potential are further modified for the inclusion of the QGP medium screening effect. This has been done through the medium dielectric permittivity, a complex quantity, which leads to a complex potential. We presented the exact numerical results and derived the analytical expression in the small velocity limit for both the real and imaginary parts of the potential. We have shown in the plots the variation of potential with respect to several parameters, such as distance between quark-antiquark, temperature, velocity, and also angular dependence. We have also presented a comparison of Coulombic and Cornell potential, considering the presence and absence of string terms. As expected, the Coulombic contribution dominates at the short distance, whereas the string term dominates at a large quarkonium separation distance. Next, it is observed that the motion of quarkonium through the QGP breaks down the spherical symmetry of the potential, and the anisotropy of the potential increases with the increase in velocity. It has also been noted that the velocity dependence of the potential is as important as the temperature dependence. The maximum variation of both the real and imaginary part of the potential from the corresponding static case is found to be along the direction of motion of the quarkonium. Finally, we obtained thermal width, which decreases with velocity and increases with temperature. This tells us that the lifetime of a quarkonium bound-state is determined by the  velocity of the quarkonium and temperature of the medium. The real part of the potential becomes positive quickly as distance increases with velocity, more prominent in the direction of quarkonium motion than perpendicular, i.e., Debye sphere shrinks and deformed.
		
		As a continuation of the present work, we would like to use the potential derived in this article to study the dynamics of the heavy quarkonia propagating in the QGP medium. The binding energy can be calculated by solving the Schr\"odinger equation using the real part of the potential. The velocity and the angular dependence of the potential are expected to modify the survival probabilities of the quarkonia. This will be a matter of investigation in the near future.

		\section{Acknowledgment}
		The authors want to thank A.~Jaiswal for the useful discussions. M.Y.J. would like to thank NISER for the financial support during his postdoctoral position and would like to acknowledge SERB-NPDF~ (National Postdoctoral Fellowship) with File No. PDF/2022/001551. N.H. is supported in part by the SERB-Mathematical Research Impact Centric Support (MATRICS) under Grant No. MTR/2021/000939.

	\end{document}